\documentclass[aps,prl,twocolumn,floatfix]{revtex4}
\usepackage{graphicx}
\usepackage{dcolumn}
\usepackage{bm}
\usepackage{latexsym}
\usepackage{amssymb}
\usepackage{amsfonts}
\usepackage{amsmath}
\usepackage{fancybox}
\usepackage{colordvi}
\usepackage{ulem}
\begin{document}
\title{Interacting drift-diffusion theory for photoexcited electron-hole gratings in semiconductor quantum wells}
\author{Ka Shen and G. Vignale}
\affiliation{Department of Physics and Astronomy, University of
Missouri, Columbia, Missouri 65211, USA} 
\begin{abstract}
{  
Phase-resolved transient grating spectroscopy in semiconductor quantum wells has been  shown to be a powerful technique for measuring  the electron-hole drag resistivity $\rho_{eh}$, which depends on the Coulomb interaction between the carriers.    In this paper we develop the interacting drift-diffusion theory,  from which  $\rho_{eh}$ can be determined, given the measured mobility of an electron-hole grating.  From this theory we predict a cross-over from a high-excitation-density regime,  in which the mobility has the ``normal" positive value, to a low-density  regime, in which Coulomb-drag dominates and the mobility becomes negative.  At the crossover point, the mobility of the grating vanishes.}  

\end{abstract}
\maketitle

The phenomenon of Coulomb drag, whereby an electronic current driven
in a quasi-two dimensional electron gas drags   carriers in an
 adjacent  quantum well, creating a potential difference or a
current in the latter has been a
topic of intense interest in condensed matter physics for the past two
decades~\cite{pogrebinskii_1977,price_physicaB_1983,zheng_prb_1993,
jauho_prb_1993,Vignale96,rojo_jpcm_1999,gramila_prl_1991,sivan_prl_1992,
Badalyan07,Croxall_prl_2008}. 
The effect was first predicted, theoretically, by 
Pogrebinskii~\cite{pogrebinskii_1977} and
Price~\cite{price_physicaB_1983}, but the field exploded only in the early
1990s, following the first successful realization of independently
contacted bilayer structures~\cite{Eisenstein1}. Since then the field has flourished, 
due to the insights it affords
on the nature of Coulomb correlations, non-equilibrium fluctuations,
and quantum coherence between spatially separated systems.  While the
original experiments were done on electronic bilayers~\cite{gramila_prl_1991}, it was soon
realized that electron-hole bilayers  offer  even more
interesting scenarios~\cite{sivan_prl_1992,Vignale96,Seamons09,
Croxall_prl_2008}.  For
example, electrons and holes can condense in an excitonic superfluid,
with or without a magnetic field~\cite{Vignale96,Balatsky04,Kellog02,Eisenstein12}, and the occurrence of such a
condensation should lead to striking manifestations in Coulomb drag
experiments~\cite{Eisenstein12}.  More recently,  drag
effects have also been investigated for different spin populations in
a single quantum well (spin Coulomb
drag)~\cite{Kikkawa99,Flatte00,scd_giovanni,scd_flensberg,weber_nature_2005,Cluj05}, in bilayer
graphene~\cite{Hwang11,Kim11}, in topological
insulators~\cite{Mink12}, and in trapped cold atoms~\cite{Sommer11}. 

The study of drag effects in electron-hole bilayers is notoriously
complicated by the fact that, in order to host carriers of opposite
polarities, the two layers must be held at different chemical
potentials, while their spatial separation is of the order of
20-30~nm.  In contrast to this, a non-equilibrium non-homogeneous
distribution of electrons and holes can be rather
easily achieved {\it in a single layer}  
with a laser of frequency larger than
the band gap, which creates an equal number of electrons and holes, in
addition to the carriers (say electrons) that are already present at
equilibrium. 

In a recent series of experiments ~\cite{Cameron96,weber_nature_2005,Yang11} 
the interference between two  laser beams coming from
different directions and polarized in the same direction has been exploited to create a {\it transient electron-hole (e-h) grating}, i.e., a spatially periodic modulation of the electron and hole densities on the surface of an $n$-type GaAs quantum well.  
Electron-hole drag  manifests itself in quite a
striking and direct way in the {\it mobility} of the e-h grating under the action of an  electric field~\cite{Hopfel86,Yang11}. 
The crucial observation is that the drift velocity of the
e-h grating - a collective formation - differs, in general,
 from the drift velocity of the background majority carriers
(electrons).  If the background carriers were absent (e.g.  in an intrinsic
material) then the drift velocity would vanish, since the net electric
force on a neutral object is zero.  In the opposite limit, in which
the e-h grating is a small perturbation on the background electron
density, the mobility depends crucially on the rate of
momentum transfer between electrons and holes~\cite{Hopfel86}.  Let us assume at first
that this can be neglected.  Then the constraint of charge neutrality
causes the dense electrons to follow the few holes rather than the other way around.  
Under these conditions the grating moves with the drift velocity of the holes, and
the mobility is positive~\cite{Smith}.  
Coulomb  collisions between
counter-flowing electrons and holes produce the equivalent of an
electric field that pushes the holes in the direction opposite to the
external  field.  Which of the two fields ultimately dominates
depends on the mobility of the electrons:  if that mobility
is sufficiently high the holes will inevitably reverse their direction of flow.
The mobility of the grating will then be negative and approach the
mobility of the electrons when the latter is  large. 

The possibility of observing anomalous (i.e., negative) mobility for an
electron-hole packet in $n$-type semiconductors was first pointed out by McLean and
Paige~\cite{McLean60} and recently confirmed by Yang {\it et
  al.}~\cite{Yang11} in $n$-GaAs.  However, these authors relied for their  analysis
on a simplified phenomenological model in which the e-h grating and the background electrons are treated as separate entities.  The e-h grating, being a neutral entity, does not directly respond to the electric field.  Hence, it is impossible, within this model,  to describe the competition between Coulomb drag and the ``normal" ambipolar mobility of the grating: the   resulting  formulas are correct only in the limit of strong Colulomb drag and  cannot be used to predict the cross-over between the ``normal" and the ``anomalous" (i.e. Coulomb drag-dominated)  regimes. 

In this paper we develop a full-fledged {interacting} drift-diffusion theory from which we derive general formulas for the mobility and the diffusion constant of the e-h grating.  These formulas not only enable us to extract precise values of the Coulomb drag
resistivity from Doppler velocimetry data, but they also lead to a striking prediction:  the e-h grating mobility can be driven through a sign reversal by changes in excitation intensity, temperature, or background density.  With an appropriate choice of parameters, the e-h grating can be made stationary in the presence of an electric field.  

Our starting point is the drift-diffusion equation for electron and hole densities ($n$ and $p$ respectively),  in which, however, we take into account the presence of the off-diagonal homogeneous conductivity $\sigma_{eh}$ (this is what makes our theory {\it interacting}, whereas, in the ``standard" theory, $\sigma_{eh}$ is set to zero).  Thus, we have
\begin{equation}
  e\partial_t 
  \begin{pmatrix}
    -n\\p
  \end{pmatrix}
  =-\nabla \cdot \left[
  \begin{pmatrix}
    \sigma_{ee} &\sigma_{eh}\\
    \sigma_{he} &\sigma_{hh}
  \end{pmatrix}
  \begin{pmatrix}
    \mathbf E_e^{\rm e}\\
    \mathbf E_h^{\rm e}
  \end{pmatrix}
\right],
  \label{eq_conth}
\end{equation}
where the effective electric fields are expressed as $\mathbf
E^{\rm e}_{e/h}=\mathbf E\pm\frac{1}{e}\nabla{\mu_{c/v}}$. Here,
$\mathbf E$ includes both external and built-in electric
fields. $\mu_c$ and $\mu_v$ are chemical potentials of electrons and
holes, respectively, not to be confused with the homogeneous mobilities $\mu_e$ and $\mu_h$, which will appear later.  By substituting the electric fields, the
continuity equations become
\begin{eqnarray}
  e\partial_t n&=& \nabla \cdot [e D_{ee}\nabla
  n+e{D_{eh}}\nabla p]+\mathbf E \cdot \nabla \sigma_e\nonumber\\ 
  &&\mbox{}  +
  \sigma_e \nabla\cdot\mathbf
  E\,,
\label{eq_fe}\\
  e\partial_t p&=&  \nabla \cdot [e D_{hh}\nabla p+eD_{he}\nabla
  n]-\mathbf E \cdot \nabla \sigma_{h}  \nonumber\\
  &&\mbox{}
  -\sigma_h \nabla\cdot\mathbf  E\,,
\label{eq_fh}
\end{eqnarray}
where $\sigma_e=\sigma_{ee}+\sigma_{eh}$ and
$\sigma_h=\sigma_{hh}+\sigma_{eh}$. The diffusion matrix elements are
$D_{\alpha\beta}=(-1)^{\delta_{\alpha\gamma}}\sigma_{\alpha\gamma}(\chi^{-1})_{\gamma\beta}/e^2$ (sum over $\gamma$ implied),
where $\chi_{\gamma\beta}=\partial n_\gamma/\partial \mu_\beta$ is the susceptibility matrix~\cite{susc}.

Following the standard procedure~\cite{Smith} we combine Eqs.\,(\ref{eq_fe}) and (\ref{eq_fh})
to cancel the space charge term $\nabla \cdot\mathbf E$, and only then impose the charge neutrality condition   $\delta p=\delta
n$, where $\delta n$ and $\delta p$ are the deviations of the corresponding densities from equilibrium and we assume ($|\nabla \delta n|/n\ll 1$).  The result is 
\begin{equation}
  \partial_t (\delta n)=-\mu_a \mathbf E_{\rm ext}\cdot\nabla \delta
  n +D_a\nabla^2 \delta n\,,
\end{equation}
where the e-h grating mobility and diffusion constant are defined by
\begin{eqnarray}
  \mu_a&=&\frac{ \left[
    \sigma_e    (\partial_n+\partial_p)\sigma_h-
    \sigma_h(\partial_n+\partial_p)\sigma_e
  \right]}{e(\sigma_e+\sigma_h)}
 ,\label{amb_mobility}\\
  D_a&=&\frac{\left(\sigma_eD_h
    +\sigma_hD_e
    \right)}{(\sigma_e+\sigma_h)}\,.
  \label{amb_diffusion}
\end{eqnarray}
Here $D_h=D_{hh}+D_{he}$ and $D_e=D_{eh}+D_{ee}$.
By taking into account the electron-hole Coulomb drag, the conductance
matrix can be expressed as~\cite{scd_giovanni} 
\begin{eqnarray}
\hat\sigma&=&
({\rho_e\rho_h+\rho_{eh}\rho_h\tfrac{\mu_h}{\mu_e}
    +\rho_{eh}\rho_e\tfrac{\mu_e}{\mu_h}})^{-1}\nonumber\\
\mbox{}&&\times
  \begin{pmatrix}
    \rho_h+\rho_{eh}\frac{\rho_h}{\rho_e}\frac{\mu_h}{\mu_e}&
    -\rho_{eh}\\
    -\rho_{eh} &
    \rho_e+\rho_{eh}\frac{\rho_e}{\rho_h}\frac{\mu_e}{\mu_h} 
  \end{pmatrix},
\label{m_conductance}
\end{eqnarray}
where the resistivities of electrons and holes are defined as $\rho_{e}=1/(n\mu_e e)$ and 
$\rho_{h}=1/(p\mu_h e)$, respectively (these resistivities can be straightforwardly measured in {\it homogeneous} d.c.  transport experiments in which the carriers are either electrons or holes). $\rho_{eh}$ stands for the cross-resistivity due to electron-hole Coulomb drag. By substituting the
conductivity matrix elements  into Eqs.\,(\ref{amb_mobility}) and
(\ref{amb_diffusion}), we obtain
\begin{eqnarray}
\nonumber
  \mu_a&=&\frac{-\mu_e(n{-p})/n}{1+(\mu_e/\mu_h)(\rho_e/\rho_{eh}){+(p/n)(\mu_e/\mu_h)}}
 \left\{
1-\frac{\rho_e}{\rho_{eh}}\right.\\ &&\hspace{-1.2cm}\times \left.\left[
  \frac{1+e\rho_{eh}({p\mu_e}+n\mu_h)
    {-enp(\mu_e+\mu_h)(\partial_n+\partial_p)\rho_{eh})} } 
  {1{+\rho_e/\rho_h}+e\rho_{eh}\mu_h(n{-p})^2/n}
\right]
\right\},\nonumber\\\label{fin_amb_mobility} \\
D_a&=&\frac{D_ep\mu_h+D_hn\mu_e+e\rho_{eh}\mu_e\mu_h(n-p)(D_hn-D_ep)}
{p\mu_h+n\mu_e+e\rho_{eh}\mu_e\mu_h(n-p)^2}.\nonumber\\\label{fin_amb_diffusion}
\end{eqnarray}
These two equations are the main results of this paper.  For
orientation we now discuss some limiting cases, taking, for definiteness, the $n$-type 
case,  i.e., $n>p$:

{\it i)} Weak Coulomb drag limit ($\sigma_{eh},\rho_{eh}\sim 0$):
\begin{eqnarray}
  \mu_a&\approx&\frac{n-p}{n/\mu_h+p/\mu_e},\\
  D_a&\approx&\frac{D_ep\mu_h+D_hn\mu_e}{p\mu_h+n\mu_e}.
\end{eqnarray}
These results agree with the well-known expressions for the
e-h grating mobility and diffusion constant in the
absence of Coulomb drag~\cite{Roosbroeck53,Smith}. 
The e-h grating mobility depends on the difference of electrons
and hole densities. In the $n$-type case, $\mu_a$ is positive: therefore, the
grating moves in the direction of the electric field. 

{\it ii)} Weak pumping regime ($\delta n \ll n$,  $\rho_h\gg\rho_e$):
\begin{eqnarray}
  \mu_a&\approx&\left(\frac{\rho_e}{\rho_{eh}}-1\right)\frac{\mu_e}{1+(\mu_e/\mu_h)
    (\rho_e/\rho_{eh})}, 
  \label{crossover}\\
  D_a&\approx&D_h\approx \frac{k_BT}{e}
  \frac{\mu_e\rho_e/\rho_{eh}}{1+(\mu_e/\mu_h)
    (\rho_e/\rho_{eh})}, \label{crossover_dif}
\end{eqnarray}
where the  minority carriers are assumed to be  non-degenerate.
Similar to Eq.\,(3) of Ref.\,\onlinecite{Hopfel86}, the expression of
the mobility here predicts a sign reversal 
of $\mu_a$ at $\rho_e=\rho_{eh}$. Specifically, $\mu_a$
becomes negative when the Coulomb drag resistivity is larger than the
ordinary resistivity of the electrons. One notices that the ratio
  $D_a/\mu_a$ in this 
case is only dependent on $\rho_e/\rho_{eh}$ and temperature, which
suggests a neat way to extract the drag resistivity in experiment~\cite{Yang11}.

{\it iii)} Strong pumping case, i.e., 
$\delta n\sim n$ and, consequently, $p\sim n$.  In this case, we find
that the expression in square bracket in Eq.\,(\ref{fin_amb_mobility})
is larger than 1 (the derivative of Coulomb drag with respect to the
density is negative), which suggests that $\mu_a$ can be positive even
for $\rho_e<\rho_{eh}$. If the carrier density due to the injection is
much larger than the one from doping, then one has $n\approx p$, resulting in
$\mu_a\approx0$.

In order to test the predictive power of Eqs.\,(\ref{fin_amb_mobility}) and (\ref{fin_amb_diffusion}) we need a reasonable model for  the Coulomb drag resistivity $\rho_{eh}$ and the homogeneous  drag-free mobilities $\mu_{e}=e/[m_{e}^\ast\sum_i(\tau^{e}_i)^{-1}]$ and $\mu_h=e/[m_{h}^\ast\sum_i(\tau^{h}_i)^{-1}]$, where $m^\ast_e$ and $m^\ast_h$ are the effective masses of electrons and holes.   
In two-dimensional polar materials  the dominant scattering mechanisms are~\cite{arora} (i)  polar-optical phonon scattering~\cite{arora}
 
\begin{eqnarray}
  \tau_{\rm OP}^{-1}=\frac{1}{4\pi\epsilon_0}\frac{e^2\pi
    \sqrt{2m^\ast\omega_0/\hbar}N_{LO}}{2\hbar}
 \left(\frac{1}{\kappa_\infty}-\frac{1}{\kappa_0}\right)\,,
\label{rateLO}
\end{eqnarray}
where the temperature dependence comes from the phonon number
$N_{LO}$,  (ii) acoustic phonon deformation potential scattering~\cite{Price81}
\begin{eqnarray}
  \tau_{D}^{-1}=
\frac{\Xi^2k_BTm^\ast}{Dv_{sl}^2\hbar^3}\frac{3}{2a},
\label{rateAC}
\end{eqnarray}
and (iii)  ionized impurity scattering, which, in the random phase approximation, has the form
\begin{eqnarray}
  \nonumber
  \tau_{\rm imp}^{-1}(\epsilon_k)&=&\frac{n_ie^4}{16\pi\hbar
    E_F\kappa_0^2\epsilon_0^2}\int_0^{2\pi} d\theta 
  (1-\cos\theta)\\
  && \mbox{}\times
\left(\tfrac{\kappa_D}{k_F}+\sqrt{{2\epsilon_k(1-\cos\theta)}/{E_F}}
  \right)^{-2},
\label{rateIM}
\end{eqnarray}
where the screening wave vector is $\kappa_D=\frac{e^2}{2\epsilon_0\kappa_0}
  \frac{1}{\hbar^2\pi}[m_e^\ast(1-e^{-\hbar^2 n\beta\pi/m_e^\ast})+m_h^\ast(1-e^{-\hbar^2 p\beta\pi/m_h^\ast})]$  and $E_F$ ($k_F$) is
the Fermi energy (wave vector).
$\kappa_0$ and $\kappa_\infty$ in these equations stand for the
dielectric function in the static and 
high frequency limits, respectively. $\epsilon_0$ is the
permittivity in vacuum and $\hbar\omega_0$ is the optical-phonon
energy. In Eq.\,(\ref{rateAC}), $D$ and
$v_{sl}$ represent the crystal density and longitudinal sound
velocity, respectively. $a$ is the effective
well-width. $\Xi$ is the deformation potential constant.
Finally, the resistivity due to electron-hole Coulomb drag is obtained from~\cite{scd_giovanni}
\begin{eqnarray}
  \rho_{eh}&=&\frac{\hbar^2}{e^2npk_BT}\frac{1}{(2\pi)^2}
  \int_0^\infty dq\frac{q^3}{2}\int_0^\infty d\omega 
  {{\rm Im}Q_{0e}{\rm Im}Q_{0h}}\nonumber\\
&&\mbox{}\times{[(1+\kappa_D/q)^2\sinh^2(\hbar\omega/2k_BT)]}^{-1},
\end{eqnarray}
where   $Q_{0e/h}$ is  the temperature-dependent Lindhard function for
electrons or holes  times the Fourier transform of the Coulomb
interaction (see Ref. \onlinecite{giuliani}). 
By taking the electron doping density 
$n_0=1.9\times 10^{11}$~cm$^{-2}$ and fitting the
low temperature electron mobility measured in the experiment by Yang
{\it et al.}~\cite{Yang11} in $9$~nm GaAs quantum
  wells,
$\mu_e(5{\rm K})=5.5\times 10^{5}$~cm$^2$/V, we determine the
effective impurity density to be $n_i=0.005n_0$. 
Unfortunately, the same model, when applied to holes, leads to a hole mobility 
about five times larger than  the one implied by the experimentally measured values of  $\mu_a$ and $D_a$. This discrepancy may be attributed to the  band structure or additional scattering mechanisms for holes.   We have found that rescaling $\mu_h$ by a factor 0.2 at all temperatures leads to results in excellent agreement with experiment.  The following calculations are based on this rescaling.

\begin{figure}
 \includegraphics[width=5.7cm]{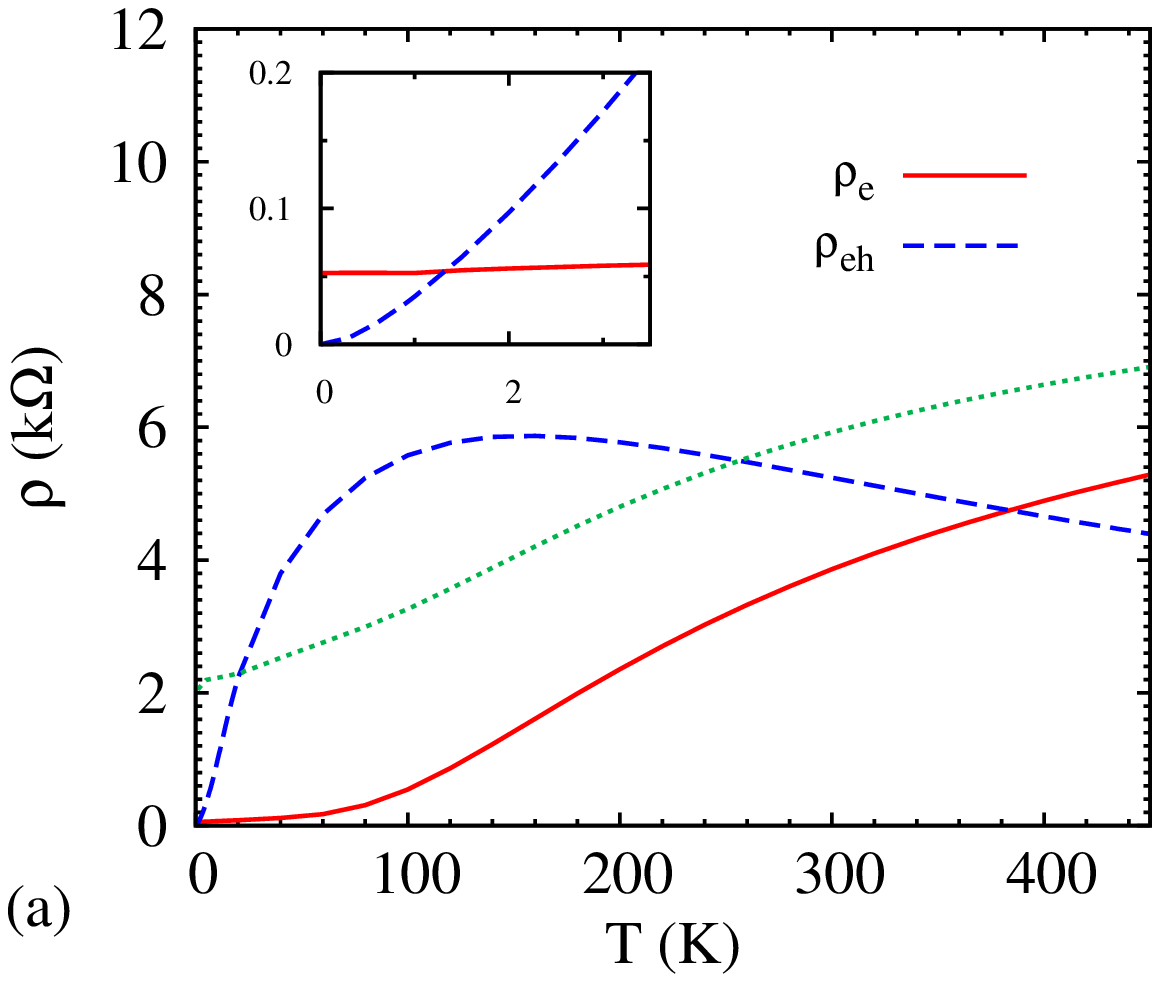}\\
  \hspace{-0.3cm} \includegraphics[width=6cm]{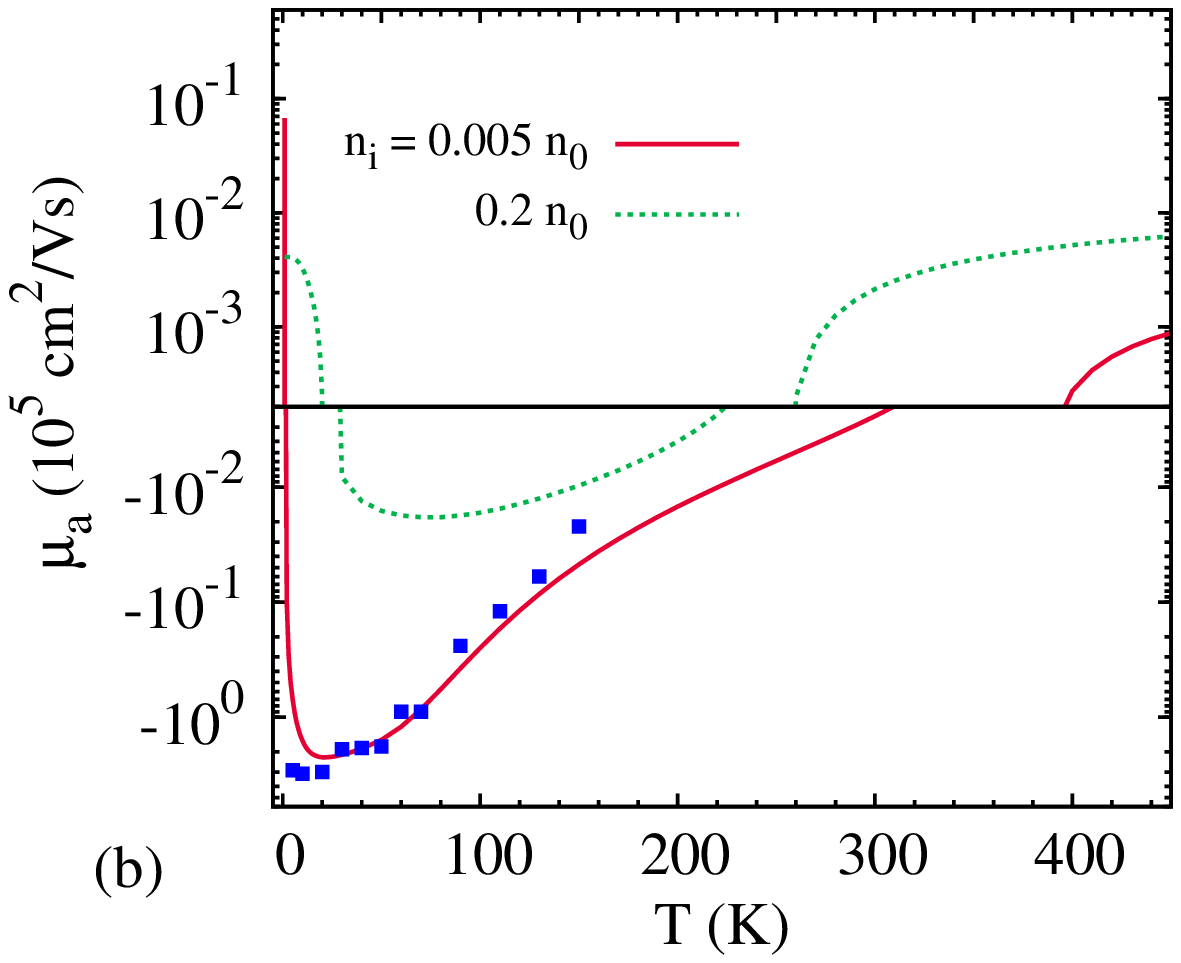}
  \caption{(Color online) (a) $\rho_e$ (red solid curve) and $\rho_{eh}$
  (blue dashed one) as functions of temperature with $n_i=0.005n_0$. The
  green dotted curve represents $\rho_e$ with $n_i=0.2n_0$. Inset: zoom in on the low temperature region $T< 3.5$ K. 
  (b) Grating mobility as function of temperature. The blue squares are experimental
  data from Ref.\,\onlinecite{Yang11}. The apparent breaks in the
  curves as they cross zero are artifacts of the logarithmic scale. 
} 
\label{fig1}
\end{figure}

In Fig.\,\ref{fig1}(a), we plot the electron resistivity
$\rho_e$ as a function of temperature, as well
as the electron-hole drag resistivity $\rho_{eh}$ in the weak pumping
regime. It is seen that the electron-hole drag resistivity is
smaller than the electron resistivity  both at high and low
temperature. However, at intermediate temperature,
one has $\rho_{eh}>\rho_e$. The two curves intersect  at 1.5~K and
385~K. For comparison, we also plot the electron resistivity (green
dotted curve) for  a more strongly disordered system, in which $n_i=0.2n_0$. 
In this case, the regime in which $\rho_{eh}>\rho_e$ is significantly
narrower, due to the increase of electron resistivity, but still clearly visible  (the
electron-hole Coulomb drag is essentially independent of impurity concentration).

Fig.\,\ref{fig1}(b) shows the spin-grating mobility $\mu_a$ calculated
from Eq.\,(\ref{fin_amb_mobility}) in the infinitesimal pumping
regime.  Two changes in sign of $\mu_a$ vs temperature are visible:
they correspond to the two crossings between $\rho_{eh}$ and $\rho_e$.    
In this figure, we also plot the experimentally determined $\mu_a$ 
(blue squares). We find that the theoretical results are in excellent agreement  with the experiment in the temperature range 20-150 K, in which range the e-h drag dominates and the e-h grating mobility is negative. 
However, the experimental data do not show any sign reversal.
On the high temperature end, it simply appears that the experiment did not reach  sufficiently high temperature.  A high-temperature cross-over was, in fact,  observed in 
Ref.~\onlinecite{Hopfel86}. 
At low temperature, the reason for the discrepancy may lie in an
undesired  laser-induced heating of the electrons.

\begin{figure}
  \includegraphics[width=5.5cm]{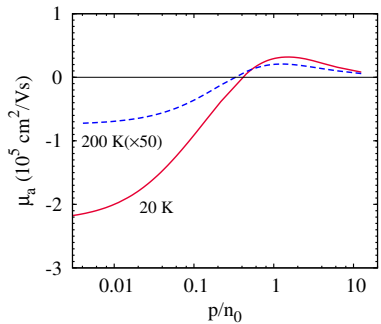}
  \caption{(Color online) Grating mobility as a function of
    pumping intensity (the latter quantified by the ratio $p/n_0$) at 20 and 200~K.} 
  \label{fig2}
\end{figure}

In Fig.\,\ref{fig2}, we plot the e-h grating  mobility as a function of
pumping intensity   at 20~K and 200~K.  Neglecting the equilibrium hole density, we have $p=\delta n$ and $n=n_0+p$.  Then, the pumping intensity can be quantified by the ratio $p/n_0$.
We find that the mobility at first increases
with increasing  pumping intensity and becomes positive at large excitation
levels. After reaching the maximum value, the mobility
begins to decrease upon further increase of the pumping intensity, and eventually approaches
zero. This behavior is consistent with our discussion above. 

\begin{figure}
  \includegraphics[width=5.5cm]{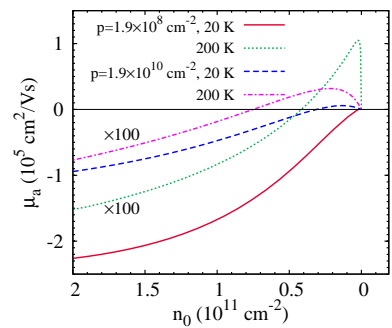}
\caption{(Color online) Grating mobility as a function of equilibrium electron
  density.
} 
\label{fig3}
\end{figure}

Another way to change the sign of the  e-h grating mobility is by
tuning the equilibrium electron density  by means of electrostatic gating. 
In Fig.\,\ref{fig3}, we show $\mu_a$ vs $n_0$ for different temperatures and
pumping intensities. In all cases, the grating mobility at first
increases with decreasing  background density and reaches a maximal 
value. This  can be understood by the quick increase of the electron
resistivity, since $\rho_e/\rho_{eh}$ increases. As the background
electron density further decreases to a negligible value, we fall back
into the ``strong pumping limit" ($n \simeq p$) and the mobility
decreases towards zero.

\begin{figure}
  \includegraphics[width=6.cm]{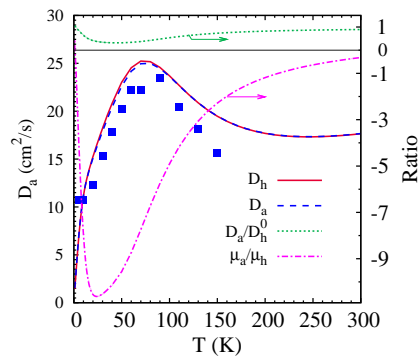}
  \caption{(Color online) Ambipolar diffusion constant (blue dashed
    curve) and hole diffusion constant (red solid curve) as function of
    temperature for $n_b=n_0$ in the weak pumping case.  Notice that the two curves are essentially identical.  The filled squares
    represent the experimental data for the ambipolar diffusion constant  reported by
    Yang {\it et al.}~\cite{Yang11}. The green dotted 
    curve shows the ratio between the ambipolar
    diffusion constants with and without Coulomb
    drag, while the pink chain curve shows the corresponding ratio for the e-h grating mobility.}
  \label{fig4}
\end{figure}

As a last point, we observe that the e-h grating diffusion constant 
remains essentially identical to the hole diffusion constant $D_h$ (see Fig.\,{\ref{fig4}}),
both being substantially reduced by Coulomb drag caused by
interaction with the background electrons. 
We notice that, as expected,
the effect of Coulomb drag on the grating diffusion constant is much
less dramatic than its effect on the mobility.    
The theoretical results show good agreement  with the experiment.

In summary, we have developed a drift-diffusion theory for the
calculation of the mobility and diffusion grating of an electron-hole
grating in a single electronic layer, taking fully into account the
effect of Coulomb drag between electrons and holes.  Our formulas show
that the mobility of an electron-hole grating is a sensitive probe of
electron-hole drag.  Further, due to its dependence of electron-hole
drag, the mobility of the grating can be driven through changes of
sign by changing temperature, excitation power, or background electron
density.   

We acknowledge support from ARO Grant No. W911NF-08-1-0317 (KS) and from NSF Grant DMR-1104788 (GV). We thank Luyi Yang for a careful reading of the manuscript and for sending us the experimental data from Ref.~\onlinecite{Yang11}.


\begin{thebibliography}{100}
\bibitem{susc} In principle, static susceptibility
  $\lim_{q=0}\chi_{\alpha\beta}(q,0)=\delta_{\alpha\beta}\chi_{\alpha}+\chi_{\alpha}\chi_\beta
  V_{q=0}$,
  where $\chi_{e}=\partial_{\mu_c} n$ and $\chi_{h}=\partial_{\mu_v}
  p$. $V_{q}$ is the screened Coulomb interaction
  element. However, we find the correction of the ambipolar diffusion constant due
  to Coulomb correlation in susceptibility is negligible in the
  present case.
\end{thebibliography}
\end{document}